\documentclass[twocolumn,aps,prd,10pt,superscriptaddress,longbibliography,floatfix,citeautoscript]{revtex4-2}

\usepackage{amsmath}
\usepackage{amssymb}
\usepackage{bm}
\usepackage{dcolumn}
\usepackage{glossaries}
\usepackage{graphicx}
\usepackage{multirow}
\usepackage{fancyvrb}
\usepackage[linesnumbered,ruled]{algorithm2e}
\usepackage{siunitx}

\usepackage[usenames,dvipsnames,svgnames]{xcolor}
\usepackage{hyperref}
\hypersetup{
    pdfnewwindow=true,      
    colorlinks=true,        
    linkcolor=Blue,         
    citecolor=Blue,         
    filecolor=Blue,         
    urlcolor=Blue           
}

\newacronym{dft}{DFT}{density functional theory}
\newacronym{emd}{EMD}{equilibrium molecular dynamics}
\newacronym{hnemd}{HNEMD}{homogeneous non-equilibrium molecular dynamics}
\newacronym{md}{MD}{molecular dynamics}
\newacronym{mlp}{MLP}{machine-learned potential}
\newacronym{nep}{NEP}{neuroevolution potential}
\newacronym{rmse}{RMSE}{root mean square error}

\begin{document}

\title{Million-atom heat transport simulations of polycrystalline graphene approaching first-principles accuracy enabled by neuroevolution potential on desktop GPUs}

\author{Xiaoye Zhou}
\affiliation{College of Physical Science and Technology, Bohai University, Jinzhou 121013, P. R. China}

\author{Yuqi Liu}
\affiliation{College of Physical Science and Technology, Bohai University, Jinzhou 121013, P. R. China}

\author{Benrui Tang}
\affiliation{College of Physical Science and Technology, Bohai University, Jinzhou 121013, P. R. China}

\author{Junyuan Wang}
\affiliation{College of Physical Science and Technology, Bohai University, Jinzhou 121013, P. R. China}

\author{Haikuan Dong}
\email{donghaikuan@163.com}
\affiliation{College of Physical Science and Technology, Bohai University, Jinzhou 121013, P. R. China}

\author{Xiaoming Xiu}
\email{xiuxiaomingdl@126.com}
\affiliation{College of Physical Science and Technology, Bohai University, Jinzhou 121013, P. R. China}

\author{Shunda Chen}
\email{phychensd@gmail.com}
\affiliation{Department of Civil and Environmental Engineering, George Washington University,
Washington, DC 20052, USA}

\author{Zheyong Fan}
\email{brucenju@gmail.com}
\affiliation{College of Physical Science and Technology, Bohai University, Jinzhou 121013, P. R. China}

\date{\today}

\begin{abstract}
First-principles molecular dynamics simulations of heat transport in systems with large-scale structural features are challenging due to their high computational cost. 
Here, using polycrystalline graphene as a case study, we demonstrate the feasibility of simulating heat transport with near first-principles accuracy in systems containing over 1.4 million atoms, achievable even with consumer desktop GPUs. 
This is enabled  by the highly efficient neuroevolution potential (NEP) approach, as implemented in the open-source GPUMD package. 
Leveraging the NEP model's accuracy and efficiency, we quantify the reduction in thermal conductivity of polycrystalline graphene due to grain boundaries with varying grain sizes, resolving contributions from in-plane and out-of-plane (flexural) phonon modes. 
Additionally, we find that grain boundaries can lead to finite thermal conductivity even under significant tensile strain, in contrast to the divergent behavior observed in pristine graphene under similar conditions, indicating that grain boundaries may play a crucial role in thermal transport in low-dimensional momentum-conserving systems.
These findings could offer insights for interpreting experimental observations, given the widespread presence of both large-scale grain boundaries and external strains in real materials.
The demonstrated ability to simulate millions of atoms with near-first-principles accuracy on consumer desktop GPUs using the NEP approach will help make large-scale high-fidelity atomistic simulations more accessible to the broader research community.
\end{abstract}

\maketitle

\section{Introduction}

Since the pioneering work by Sosso \textit{et al.} \cite{sosso2012prb}, which employed the Behler-Parrinello neural network potential \cite{behler2007prl} to study heat transport in phase-change materials, \glspl{mlp}  have gradually gained popularity in studying heat transport in various materials \cite{dong2024jap}.
However, most previous works have focused on relatively simple systems, such as three-dimensional crystals \cite{cheng2023lattice, du2023low, ying2023sub, wang2024dissimilar, Zeraati2024prm}, two-dimensional and layered materials \cite{dong2023ijhmt, sha2023phonon, eriksson2023tuning, Li2024prb, Li2024ijhmt_Active}, amorphous materials \cite{wang2023prb, zhang2023prb, liang2023prb,li2024ijhmt, Pegolo2024FM}, simple liquids \cite{Tisi2021PRB, xu2023jcp}, and heterojunctions \cite{Mangold2020jap, wu2024nc}. These systems typically involve relatively small structural features and are computationally manageable with \glspl{mlp}. 
To date, no applications of \glspl{mlp} have been reported for studying heat transport in large-scale polycrystalline systems, mainly due to the higher computational cost of most \glspl{mlp} compared to traditional empirical potentials. 
In this work, we address this gap by demonstrating the feasibility of \glspl{mlp} in studying heat transport in large-scale polycrystalline graphene systems, containing up to \num{1 438 236} atoms, which can be simulated even on consumer desktop GPUs. 
This is enabled by the highly efficient \gls{nep} approach \cite{fan2021neuroevolution, fan2022jpcm, fan2022jcp} as implemented in the open-source GPUMD package \cite{fan2017cpc}. 

Several \glspl{mlp} have been developed for carbon-based systems \cite{qamar2023jctc, wang2022cm, shaidu2021npjcm, Rowe2020jcp, fan2024jpcm}, exploring various properties such as thermal and mechanical behavior.
Here, we utilize our previously developed general-purpose \gls{nep} model for carbon systems \cite{fan2024jpcm}, which has already been applied to study thermal and thermoelectric transport in a graphene anti-dot lattice with \num{187 200} atoms. 
First, we benchmark the computational performance of this \gls{nep} model compared to typical \glspl{mlp} for carbon systems. 
Next, we validate the accuracy of the \gls{nep} model by predicting the grain boundary energies for various tilt angles. 
Finally, we apply this \gls{nep} model in \gls{md} simulations to investigate heat transport in polycrystalline graphene with different grain sizes, quantifying the reduction and the scaling of the thermal conductivity with respect to the average grain size. We also resolve the contributions from the in-plane and out-of-plane (flexural) phonon modes. Additionally, we investigate the effects of external tensile strain on the thermal conductivity of polycrystalline graphene, finding that the presence of grain boundaries could effectively suppress the thermal conductivity divergence as observed in pristine graphene under similar conditions \cite{Pereira2013prb,fan2017prb,Xu2023ps}. This suggests that grain boundaries, along with asymmetric interatomic interactions \cite{ZhongPRE2012,chen2012break,chen2013asymmetric,ChenPRE2015LJ,chen2016key} and nearly integrable dynamics \cite{Chen2014PRE,ZhaoPRE2018,LepriPRL2020}, may play an important role in thermal transport in low-dimensional momentum-conserving  systems \cite{BONETTO2000MP,LEPRI2003PR,Dhar2008AdvPhys,Benenti2023review}. These findings could provide valuable insights for interpreting experimental observations in real materials that feature both large-scale grain boundaries and external strains. This work demonstrates that the NEP approach enables highly efficient million-atom atomistic simulations approaching first-principles accuracy, making such large-scale studies more accessible to the broader research community using either high-performance computing resources or consumer desktop GPUs.

\section{Models and Methods}

\subsection{Simulation models for polycrystalline graphene}

\begin{figure*}[htb]
\centering
\includegraphics[width=2\columnwidth]{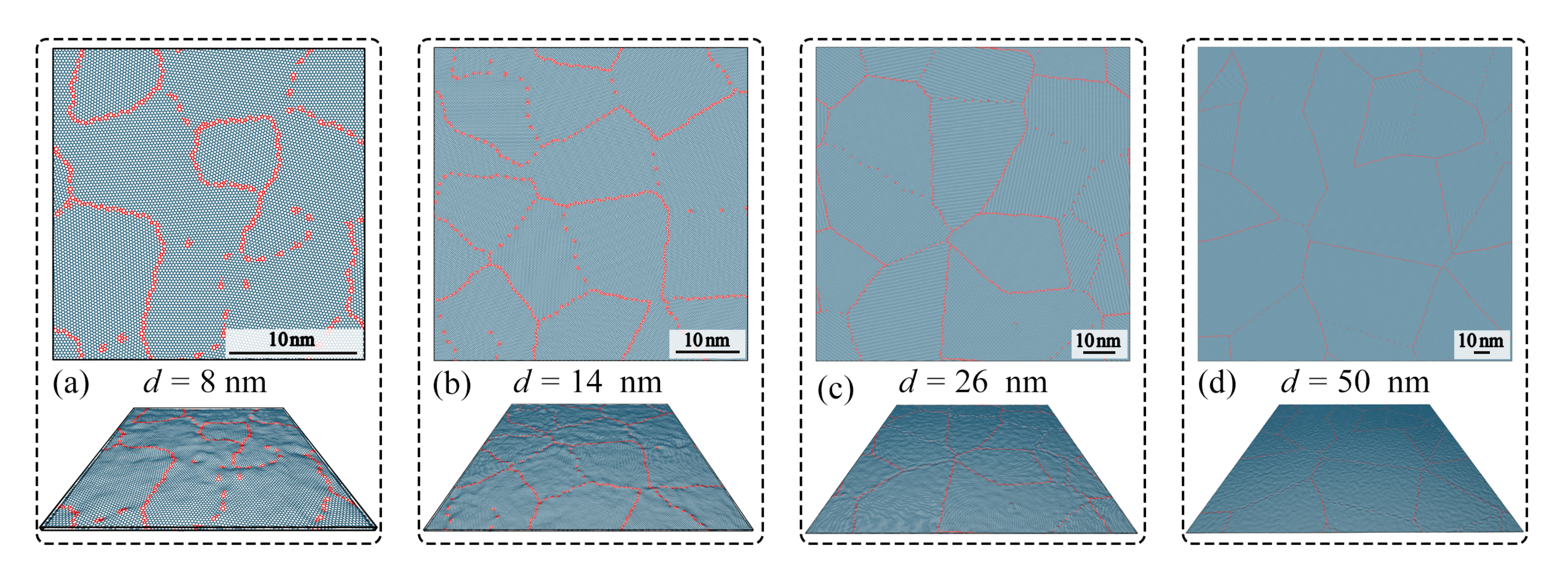}
\caption{
    Snapshots of polycrystalline graphene samples with four different grain sizes, $d$ (where $d = (A/n)^{1/2}$, with $A$ as the total graphene plane area and $n$ as the number of grains), after MD relaxation at 300 K and zero pressure: (a) $d = 8$ nm (\num{22464} atoms), (b) $d = 14$  nm (\num{89858} atoms), (c) $d = 26$  nm (\num{359502} atoms), and (d) $d = 50$ nm (\num{1438236} atoms). 
}
\label{fig:model}
\end{figure*}

We used the polycrystalline graphene samples from ~Ref. \citenum{fan2017nl}, constructed using a phase-field crystal model \cite{hirvonen2016prb}. 
The grain size, $d$, is defined as $d = (A/n)^{1/2}$, where $A$ is the total area of the graphene plane and $n$ is the number of grains comprising the plane. We considered four system sizes: \num{22464} atoms ($A = 24.3 \times 24.3$ nm$^2$), \num{89858} atoms ($A = 48.5 \times 48.5$ nm$^2$), \num{359502} atoms ($A = 96.9 \times 97.0$ nm$^2$), and \num{1438236} atoms ($A = 193.8 \times 193.9$ nm$^2$). Each sample was initially prepared with 16 randomly placed and oriented crystallites; however, after relaxation, the final number of grains was typically smaller than 16. For each system size, we prepared several independent samples with varying grain distributions: 7, 4, 2, and 3 samples,  averaging over which yielded four effective grain sizes of 8 nm ($A = 24.3 \times 24.3$ nm$^2$), 14 nm ($A = 48.5 \times 48.5$ nm$^2$), 26 nm ($A = 96.9 \times 97.0$ nm$^2$), and 50 nm ($A = 193.8 \times 193.9$ nm$^2$), respectively. 
Figure \ref{fig:model} illustrates the polycrystalline graphene samples with four different grain sizes $d$ after MD relaxation at 300 K and zero pressure.

\subsection{The NEP model for carbon systems}

In this work, we employ a \gls{nep} model constructed for general carbon systems \cite{fan2024jpcm} to describe the interatomic forces in both pristine and polycrystalline graphene. 
The \gls{nep} approach \cite{fan2021neuroevolution, fan2022jpcm, fan2022jcp} provides highly accurate and efficient \glspl{mlp} for general systems by training a neural network model using an evolutionary algorithm, which is a powerful black-box optimizer for real-valued optimization problems. 
The training process is guided by a loss function, which is a weighted sum of the \glspl{rmse} of energy, force, and virial, alongside regularization terms that help to prevent or alleviate overfitting. 
The \gls{rmse} of a physical quantity is calculated from the target values and predicted ones for the structures in the training data set.

A neural network model is employed  to predict the site energy $U_i$ of an atom in a structure. 
The energy of a structure is then calculated as a sum of the site energies of the atoms within the structure. 
As a typical many-body potential, the force and virial can be calculated straightforwardly \cite{fan2015prb}.
The input of the neural network model is a descriptor vector which consists of a number of radial and angular components. 
Both the radial and angular components involve a number of radial functions, each expressed as a linear combination of a number of basis functions based on Chebyshev polynomials. 
The target values in the training dataset were calculated based on quantum-mechanical \gls{dft} calculations, as contributed by Rowe \textit{et al.} in Ref. \citenum{Rowe2020jcp}. 
For details on the various hyperparameters used to construct the \gls{nep} model based on this training data set, we refer to Ref. \citenum{fan2024jpcm}.
\subsection{Methods for calculating the thermal conductivity}

We use both the \gls{emd} and \gls{hnemd} method sto compute the thermal conductivity. 

In the \gls{emd} method, on equilibrium state  is first achieved,and then che running thermal conductivity $\kappa_{\mu\nu}(t)$ a($\mu, \nu = x, y, z$) is calculated s a function of the correlation time $t$ according to the Green-Kubo relation \cite{green1954jcp,kubo1957jpsj}:
\begin{equation}
\label{equation:gk}
\kappa(_{\mu\nu}t) = \frac{1}{k_{\rm B}T^2V}\int_0^{t} \langle J_{\mu}(0) J_{\nu}(t') \rangle_{\rm e} dt'.
\end{equation}
Here $k_{\rm B}$ is the Boltzmann constant, $T$ is the temperature, $V$ is the volume, and $\langle J_{\mu}(0) J_{\nu}(t') \rangle_{\rm e}$ is the heat current autocorrelation function as evaluated in an equilibrium ensemble (hence the subscript ``e'').
The heat current $J_{\mu}$ can be calculated based on the general formulation for many-body potentials \cite{fan2015prb}:
\begin{equation}
\mathbf{J} = \sum_i \sum_{j\neq i} \mathbf{r}_{ij} \frac{\partial U_j}{\partial \mathbf{r}_{ji}} \cdot \mathbf{v}_i,
\end{equation}
where $\mathbf{r}_{ij}\equiv \mathbf{r}_{j} - \mathbf{r}_{i}$, $\mathbf{r}_{i}$ is the position of atom $i$, and $\mathbf{v}_{i}$ is the velocity of atom $i$.

In the \gls{hnemd} method, an external driving force is first applied for each atom, and then the running thermal conductivity is calculated based on the following linear response relation \cite{evans1982pla, fan2019prb}
\begin{equation}
\label{equation:hnemd_kappa}
\langle J_{\mu}(t) \rangle_{\rm ne} = TV \sum_{\nu} \kappa_{\mu\nu}(t) F^{\nu}_{\rm e},
\end{equation}
where $F^{\nu}_{\rm e}$ is the driving force parameter. 
For general many-body potentials, the driving force on atom $i$  is formulated as \cite{fan2019prb}:
\begin{equation}
    \mathbf{F}^{\rm ext}_i =  \mathbf{F}_{\rm e} \cdot
    \sum_{j\neq i} \mathbf{r}_{ij} \otimes 
        \frac{\partial U_j}{\partial \mathbf{r}_{ji}}.
\end{equation}
This driving force directs more energetic atoms towards the heat flow direction and less energetic ones towards the opposite direction.

We used the GPUMD package \cite{fan2017cpc} to perform all the \gls{md} simulations. 
A time step for integration of 1 fs was used throughout. In this work, we only considered the temperature of 300 K.  
In the \gls{hnemd} simulations, the magnitude of the driving force parameter was set to $F_{\rm e}=0.1$ $\rm\mu m^{-1}$ for pristine graphene and $F_{\rm e}=1$ $\rm\mu m^{-1}$  for polycrystalline graphene. 
To determine the volumes of the simulated systems, the layer thicknesses for both pristine graphene and polycrystalline graphene were set to $3.35$ \AA{}, corresponding to the equilibrium interlayer distance in graphite. 

\section{Results and Discussion}

\subsection{Performance evaluation}

\begin{figure}[htb]
\centering
\includegraphics[width=1\columnwidth]{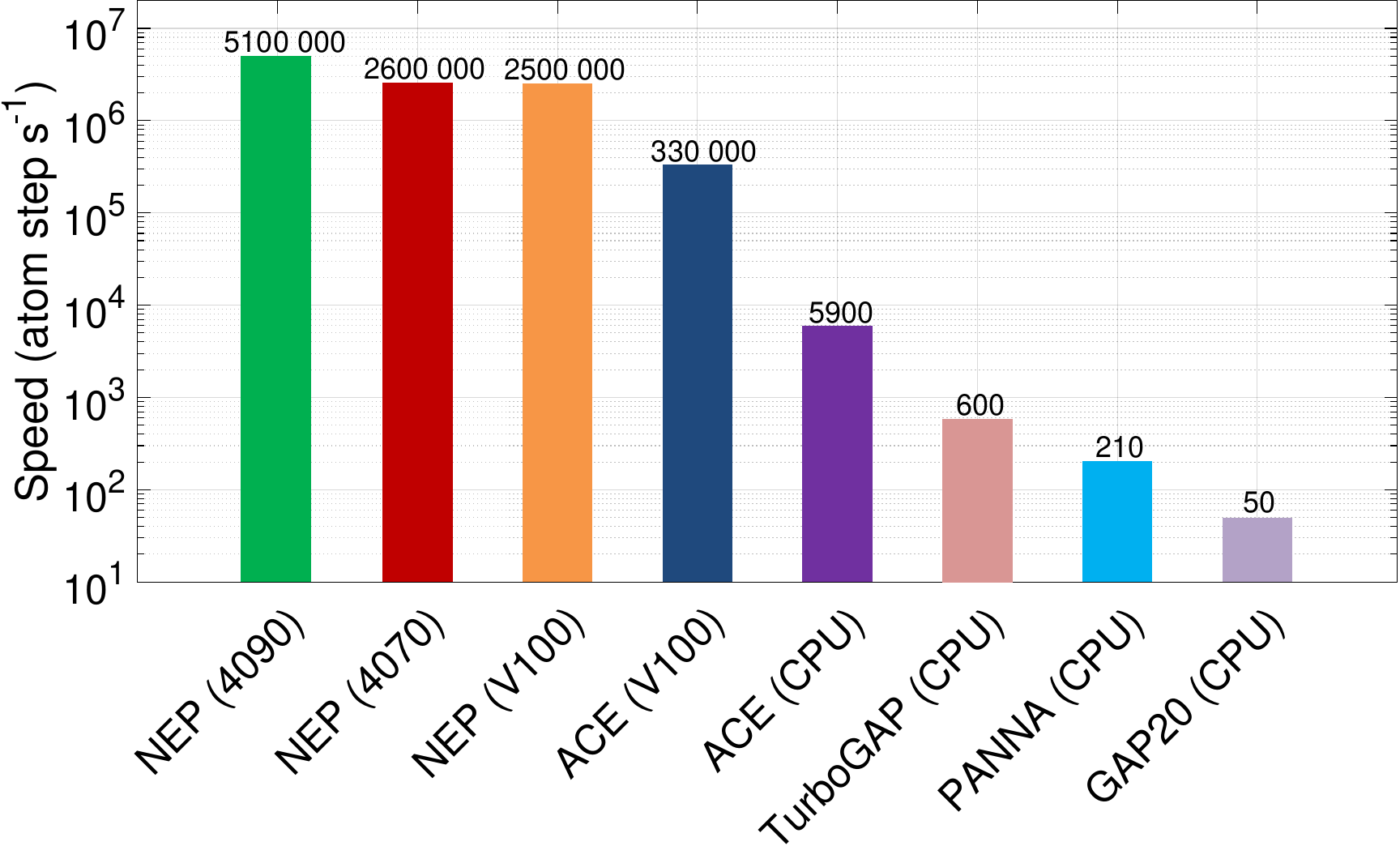}
\caption{
Comparison of the computational speeds between NEP model and typical existing MLPs for carbon systems, including ACE \cite{qamar2023jctc}, TurboGAP \cite{wang2022cm}, PANNA \cite{shaidu2021npjcm}, and GAP20 \cite{Rowe2020jcp}. Data for MLPs other than NEP are sourced from Ref. \citenum{qamar2023jctc}. NEP outperforms the previously top-performing model, ACE, by 8x on the same V100 GPU and achieves high computational performance on consumer desktop GPUs, matching V100 speed with an RTX 4070 and delivering further 2x speedup with an RTX 4090.
}
\label{fig:speed}
\end{figure}

We first evaluate the computational speed of the \gls{nep} model and compare it to several typical existing \glspl{mlp} for carbon systems, including ACE \cite{qamar2023jctc}, TurboGAP \cite{wang2022cm}, PANNA \cite{shaidu2021npjcm}, and GAP20 \cite{Rowe2020jcp}. Data for the other \glspl{mlp} were taken from Ref. \citenum{qamar2023jctc}. To ensure consistency with previous benchmarks \cite{qamar2023jctc,wang2022cm}, we performed \gls{md} simulation for liquid carbon at 4000 K with a density of 2 g/cm$^{3}$. Computational speed was measured as the product of the number of atoms and the number of simulation steps divided by the computation time. Most of the previous \glspl{mlp} were benchmarked on CPU, with the ACE model showing the highest speed \cite{qamar2023jctc}. 
The ACE model was also benchmarked on a V100 GPU. 
Our \gls{nep} model demonstrates significantly higher computational speed than the ACE model, performing about 8 times faster than ACE on the same V100 GPU. Moreover, our \gls{nep} model achieves high computational performance on consumer desktop GPUs, matching the speed of V100 with an RTX 4070 and offering a two-fold speedup with an RTX 4090.

\subsection{Grain boundary energy}

\begin{figure}[htb]
\centering
\includegraphics[width=1\columnwidth]{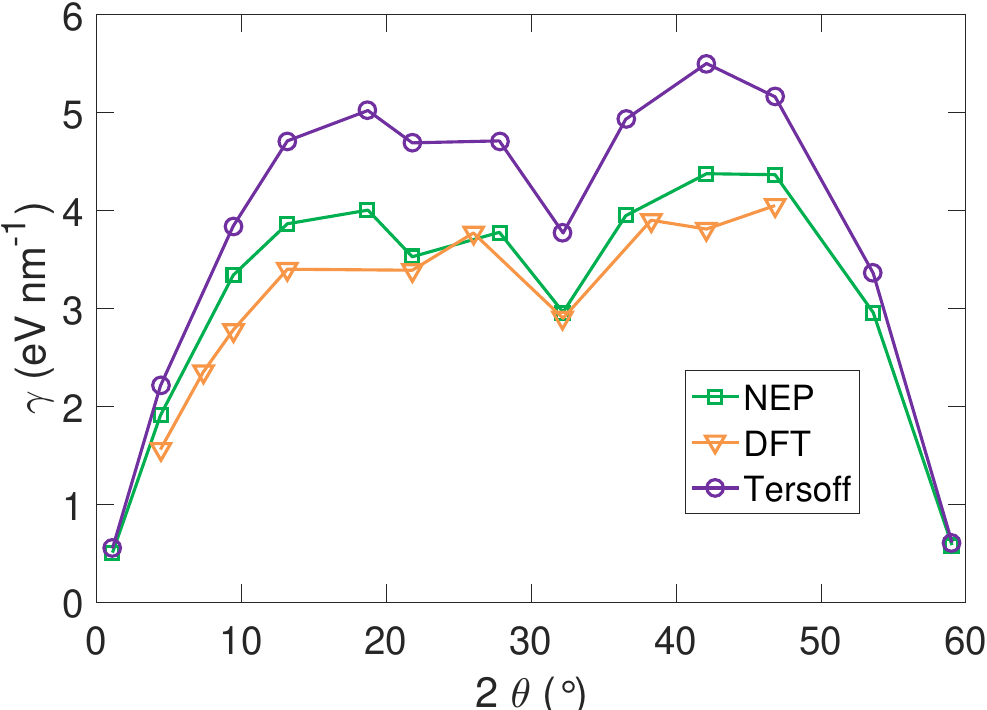}
\caption{
    Grain boundary energy density $\gamma$ as a function of tilt angle 2$\theta$, calculated using the NEP model (this work), DFT \cite{hirvonen2016prb}, and the Tersoff potential \cite{azizi2017cb}. The NEP model aligns more closely with DFT calculations.
}
\label{fig:grain_energy}
\end{figure}

Before applying the \gls{nep} model to heat transport calculations, we examine the grain boundary energies, which have a strong correlation with heat transport properties \cite{azizi2017cb, dong2018pccp}. 
To this end, we use the bicrystalline graphene samples \cite{azizi2017cb} previously constructed using the PFC method \cite{hirvonen2016prb}.
The grain boundary energy per unit length, or line tension $\gamma$, can be calculated as follows:
\begin{equation}
\gamma = \frac{E - N E_0}{2L},
\end{equation}
where $E$ is the total energy of the bicrystalline graphene system with $N$ atoms and two symmetrically tilted grain boundaries with length $L$, and $E_0$ is the energy per atom in pristine graphene.
Figure \ref{fig:grain_energy} shows the grain boundary energy density $\gamma$ as a function of the tilt angle $2\theta$ calculated using the \gls{nep} model, compared to previous results from \gls{dft} \cite{hirvonen2016prb} and a Tersoff potential \cite{azizi2017cb}.
The results from the \gls{nep} model show much better agreement with \gls{dft} calculations than the results from the Tersoff potential.

\subsection{Thermal conductivity}

After validating the \gls{nep} model's accuracy in describing the grain boundary energetics, we proceed to study the effects of grain boundary on heat transport. 
To this end, we employ the efficient \gls{hnemd} method to calculate the thermal conductivity of polycrystalline samples, closely comparing the results with those of pristine graphene. 

\begin{figure}
\centering
\includegraphics[width=\columnwidth]{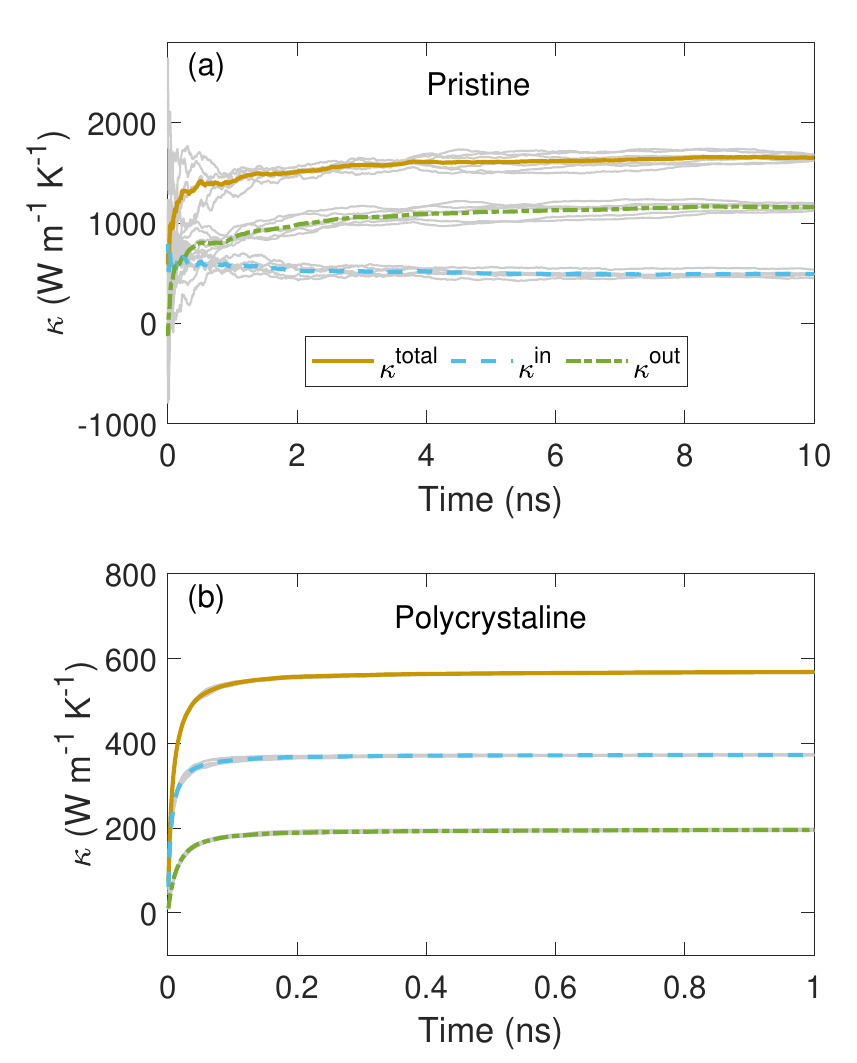}
\caption{
Cumulative average of the running thermal conductivity, showing total ($\kappa^{\rm tot}$, orange solid line), in-plane ($\kappa^{\rm in}$, blue dashed line), and out-of-plane ($\kappa^{\rm out}$, green dashed-dotted line) phonon contributions as a function of production time, calculated using the HNEMD method for (a) pristine graphene (simulation domain size: about 25 nm $\times$ 25 nm, \num{24000} atoms) and (b) polycrystalline graphene (grain size: $d = 50$ nm, \num{1438236} atoms) at 300 K and zero pressure. The thin lines represent independent runs and the thick lines show the averaged results. Six independent runs were conducted for both pristine and polycrystalline graphene.}
\label{fig:hnemd}
\end{figure}

Figure~\ref{fig:hnemd}(a) presents the cumulative average of the running thermal conductivity $\kappa(t)$, as defined in Eq.~(\ref{equation:hnemd_kappa}), for pristine graphene. 
To facilitate the discussion, the total thermal conductivity $\kappa^{\rm tot}$ is decomposed into contributions from in-plane phonons ($
\kappa^{\rm in}$) and out-of-plane phonons ($
\kappa^{\rm out}$) \cite{fan2017prb}.
For both components, the cumulative average of the running thermal conductivity converges well at a production time of 10 ns.
Upon convergence, the values for pristine graphenen are $\kappa^{\rm in} = 492 \pm 11$ W m$^{-1}$ K$^{-1}$, $\kappa^{\rm out} = 1162 \pm 12$ W m$^{-1}$ K$^{-1}$, and $\kappa^{\rm tot} = 1654 \pm 23$ W m$^{-1}$ K$^{-1}$ respectively.
This indicates that the thermal conductivity of graphene is mainly contributed by the out-of-plane phonons, i.e., the flexural modes ($\kappa^{\rm out}$). 
It has been observed \cite{wu2024jcp} that NEP tends to slightly underestimate the thermal conductivity of typical materials, including pristine graphene, due to force errors, though this underestimation could be corrected systematically.
However, since the focus of this work is not on the absolute values of thermal conductivity, we do not apply such corrections here. 

\begin{figure}
\centering
\includegraphics[width=\columnwidth]{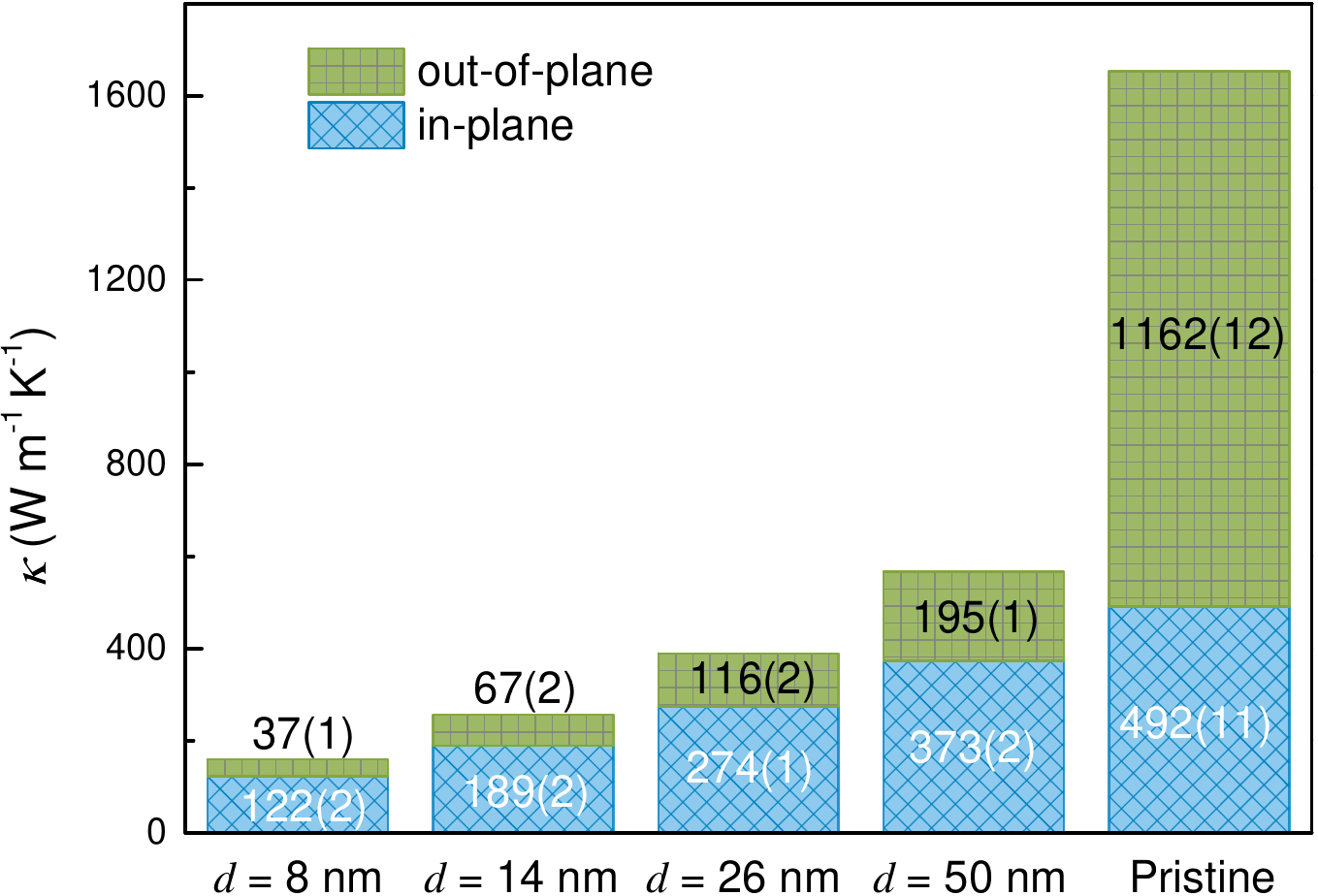}
\caption{
In-plane ($\kappa^{\rm in}$, lower blue bars), out-of-plane ($\kappa^{\rm out}$, upper green bars), and total ($\kappa^{\rm tot}$, combined blue and green bars) thermal conductivity for polycrystalline graphene with various grain sizes ($d = 8$ nm, \num{22464} atoms; $d = 14$ nm, \num{89858} atoms; $d = 26$ nm, \num{359502} atoms; $d = 50$ nm, \num{1438236} atoms) and pristine graphene (simulation domain size: approximately 25 nm $\times$ 25 nm, \num{24000} atoms) at 300 K and zero pressure.}
\label{fig:kappa_bar}
\end{figure}

Figure~\ref{fig:hnemd}(b) presents the results for the polycrystalline graphene with a grain size of $d=50$ nm.
The cumulative average of the running thermal conductivity converges well at a production time of 1 ns.
Upon convergence, the values for polycrystalline graphene are $\kappa^{\rm in} = 373 \pm 2$ W m$^{-1}$ K$^{-1}$, $\kappa^{\rm out} = 195 \pm 1$ W m$^{-1}$ K$^{-1}$, and $\kappa^{\rm tot} = 568 \pm 3$ W m$^{-1}$ K$^{-1}$ respectively.
In contrast to pristine graphene, the thermal conductivity of polycrystalline graphene is mainly contributed by the in-plane phonons ($\kappa^{\rm in}$). 
Additional results for polycrystalline graphene with other grain sizes are shown in Fig.~\ref{fig:kappa_bar}. 
As the grain size decreases, both the absolute and relative contributions of the flexural phonons to the total thermal conductivity decrease.

\begin{figure}
\centering
\includegraphics[width=\columnwidth]{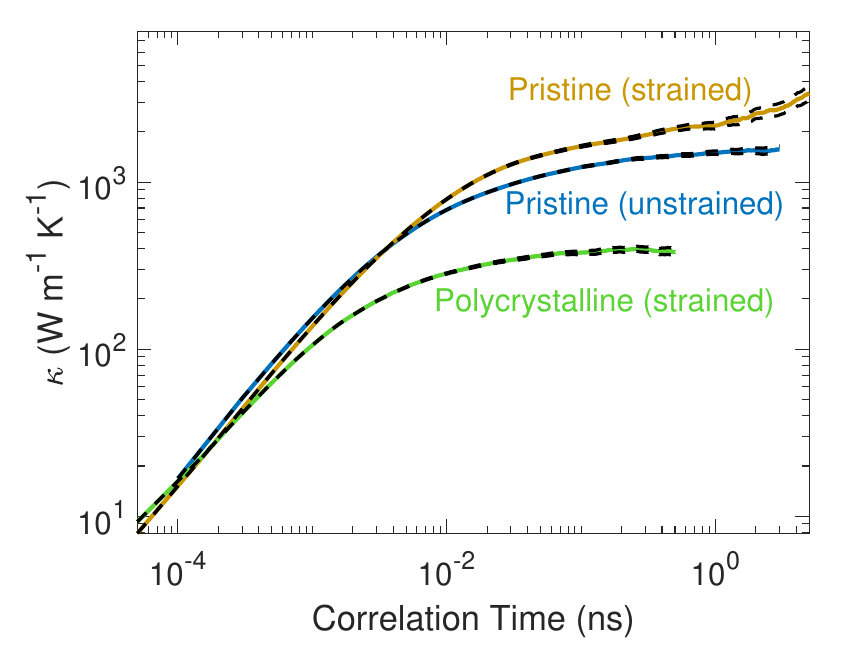}
\caption{
Running thermal conductivity as a function of correlation time for pristine graphene under biaxial tensile strain (40 GPa stress, orange line) and without strain (blue line), and polycrystalline graphene (grain size $d=14$ nm) under the same biaxial tensile strain (40 GPa stress, green line) at 300 K. Black dashed lines represent the error bounds calculated from the standard error based on about 100 independent \gls{emd} runs. For pristine graphene, the simulation domain size under zero pressure is about 25 nm $\times$ 25 nm. Unlike pristine graphene, polycrystalline graphene under biaxial tensile strain (40 GPa stress) at 300 K does not exhibit divergent thermal conductivity.}
\label{fig_stress}
\end{figure} 

The aforementioned results are for unstrained systems. 
Next, we consider the effects of external tensile strain, which is usually unavoidable and important for two-dimensional materials. 
Previous works \cite{Pereira2013prb, fan2017prb, Xu2023ps} suggest that the thermal conductivity in sufficiently stretched graphene might be divergent, i.e., it increases with sample size (proportional to the correlation time in \gls{emd}) without an upper limit. 
This behavior is confirmed here for pristine graphene under 40 GPa biaxial tensile stress (equivalent to about 5\% biaxial tensile strain), as shown in Fig.~\ref{fig_stress}. 

For the strained systems, the \gls{emd} method is used, as the running thermal conductivity in \gls{hnemd} simulations may also show divergence in the non-linear response regime due to an extensively large driving force parameter \cite{fan2019prb}, complicating the analysis.
In strained pristine graphene, the running thermal conductivity surpasses that of unstrained pristine graphene at a correlation time of about 0.01 ns and rapidly increases after a correlation time of 3 ns, showing no signs of convergence up to a correlation time of 5 ns.
While our results do not definitely confirm the divergence of the thermal conductivity in strained pristine graphene at infinite correlation time, they are consistent with previous studies \cite{Pereira2013prb, fan2017prb, Xu2023ps}. 

In contrast, polycrystalline graphene with grain size $d=14$ nm shows well-converged thermal conductivity with respect to the correlation time under a 40 GPa biaxial tensile stress, with an enhanced convergent thermal conductivity of $399 \pm 21$ W m$^{-1}$ K$^{-1}$, compared to $256 \pm 4$ W m$^{-1}$ K$^{-1}$ under zero stress, indicating that tensile strain can enhance the thermal conductivity in both pristine and polycrystalline graphene, except in small samples in the ballistic regime \cite{Wei2011nt}.

The finite (non-divergent) thermal conductivity in polycrystalline graphene under significant tensile strain is intriguing, as it indicates that grain boundaries, along with asymmetric interatomic interactions \cite{ZhongPRE2012,chen2012break,chen2013asymmetric,ChenPRE2015LJ,chen2016key} and nearly integrable dynamics \cite{Chen2014PRE,ZhaoPRE2018,LepriPRL2020}, may play a crucial role in thermal transport in momentum-conserving low-dimensional systems, where the foundations of the empirical Fourier's heat conduction law remain to be rigorously established \cite{BONETTO2000MP,LEPRI2003PR,Dhar2008AdvPhys,Benenti2023review}. Further investigations are required to better understand the role of grain boundaries and explore the underlying mechanisms that could lead to Fourier's heat conduction law in low-dimensional momentum-conserving systems.  

\section{Summary and conclusions}

In this work, we employed the highly efficient NEP model, as implemented in the GPUMD package, to investigate heat transport in million-atom polycrystalline graphene, in close comparison with pristine graphene. 

First, we demonstrated that the NEP model achieves computational speed at least an order of magnitude faster than state-of-the-art \glspl{mlp} in the literature. 
The high efficiency of NEP makes fully atomistic simulation of heat transport in million-atom systems feasible. 

Next, we validated that the NEP model provides significantly more accurate descriptions of grain boundary energies compared to the Tersoff potential, closely matching the results from DFT calculations. 
Following this validation, we calculated the thermal conductivity of pristine and polycrystalline graphene systems using the HNEMD and EMD methods.

Our results show that the grain boundaries significantly reduce the thermal conductivity of graphene, especially limiting the contribution from the out-of-plane (flexural) phonon modes.
Moreover, we investigated the effects of tensile strain and found that it enhances the thermal conductivity of both pristine and polycrystalline graphene systems. 

However, a key qualitative difference emerges: while the thermal conductivity in sufficiently stretched pristine graphene appears to diverge with increasing correlation time (or sample size), the thermal conductivity in similarly stretched polycrystalline graphene converges to a finite value. This suggests that grain boundaries play an important role in modulating heat transport in polycrystalline graphene under strains. It also indicates that grain boundaries, along with asymmetric interatomic interactions \cite{ZhongPRE2012,chen2012break,chen2013asymmetric,ChenPRE2015LJ,chen2016key} and nearly integrable dynamics \cite{Chen2014PRE,ZhaoPRE2018,LepriPRL2020}, may be crucial in thermal transport in low-dimensional momentum-conserving  systems \cite{BONETTO2000MP,LEPRI2003PR,Dhar2008AdvPhys,Benenti2023review}. Further investigations in separate works are needed to further elucidate the role of grain boundaries and investigate the underlying mechanisms that could lead to Fourier's heat conduction law in low-dimensional momentum-conserving systems.

Our highly efficient and accurate NEP approach enables detailed investigations of large-scale systems, providing deeper insights into heat transport in graphene with grain boundaries and external strains. These findings will facilitate closer comparisons with experimental results, considering the ubiquitous presence of both large-scale grain boundaries and external strains in real-world materials. The capability of simulating millions of atoms with near-first-principles accuracy on consumer desktop GPUs using the NEP approach will enhance the accessibility to large-scale and accurate atomistic simulations for the broader research community, leveraging both high-performance computing resources and consumer-grade hardware.

\vspace{0.5cm}

\begin{acknowledgments}
H.D. is supported by the Science Foundation from the Education Department of Liaoning Province (No. JYTMS20231613) and the Doctoral start-up Fund of Bohai University (No. 0523bs008). X.Z. is supported by the College Students' Innovation and Entrepreneurship Training Program of Liaoning Province (No. S202410167146).
\end{acknowledgments}

\vspace{0.5cm}
\noindent{\textbf{Data availability:}}

Complete input and output files for the general-purpose carbon \gls{nep} model are freely available at \url{https://gitlab.com/brucefan1983/nep-data}.

\vspace{0.5cm}
\noindent{\textbf{Declaration of competing interest:}}

The authors declare that they have no competing interests.

\bibliography{refs}

\end{document}